\documentclass[aps,prl,amsfonts, amssymb, amsmath, showkeys, nofootinbib, nobibnotes, reprint]{revtex4-2}
\usepackage[utf8]{inputenc}
\usepackage[english]{babel}
\usepackage{amsmath}
\usepackage{graphicx}
\usepackage{mathtools}
\usepackage{multirow}
\usepackage{multirow}
\usepackage{float}
\usepackage{appendix}

\usepackage[colorinlistoftodos, color=green!40, prependcaption]{todonotes}

\usepackage{color}
\usepackage{ulem}

\begin{document}

\title{Phase-Flow Topology of Bound States in the Continuum}

\author{Nikolai A. Vlasov}%
\email{nikolai.vlasov@metalab.ifmo.ru}
\affiliation{School of Physics and Engineering,
ITMO University, St. Petersburg, Russia 197101}%

\author{Varvara P. Panurchenko}%
\affiliation{School of Physics and Engineering,
ITMO University, St. Petersburg, Russia 197101}%

\author{Ravshanjon Kh. Nazarov}%
\affiliation{School of Physics and Engineering,
ITMO University, St. Petersburg, Russia 197101}%

\author{Stanislav S. Baturin}%
\affiliation{School of Physics and Engineering,
ITMO University, St. Petersburg, Russia 197101}%

\author{Ekaterina E. Maslova}%
\affiliation{School of Physics and Engineering,
ITMO University, St. Petersburg, Russia 197101}%

\author{Zarina F. Kondratenko}%
\affiliation{School of Physics and Engineering,
ITMO University, St. Petersburg, Russia 197101}%


\date{\today} 

\begin{abstract}
Control of topological charges of bound states in the continuum (BICs) is essential for advanced topological photonics. A rigorous theoretical framework for understanding the formation of these charges is therefore necessary for further progress in this field. However, conventional multipolar formalism often fail to predict the topological charge when no single multipole dominates the mode. To address this fundamental issue, we present a rigorous dynamical systems framework for describing formation of topological charges. This attitude provides a direct link between polarization vortex and the local structure of the vector polarization field. Within the proposed framework, we challenge the established identification of the topological charge with the Hopf index of the dominant multipole and demonstrate the significant role played by derivatives of the multipolar coefficients in determining the charge. Our theoretical framework is validated within semi-analytical multipolar decompositions and full-wave numerical simulations of periodic dielectric metasurfaces. Furthermore, we show how the symmetry of the unit cell influences the local structure of the polarization field around BIC and identify the necessary conditions for the formation of high topological charges. Our findings provide a more rigorous basis for analysing BICs' topological properties, paving the way for advanced applications in topological photonics.

\end{abstract}

\keywords{bound states in the continuum, topological charge, vector spherical harmonics}

\maketitle

The concept of topological charge describes
the winding of a phase or polarization field around a singular point as a topological invariant that remains robust against small perturbations of the system \cite{gbur2016singular, Padgett2009, Shen, Liu2021, Jian2018, Jianlong2020}. In photonics, non-radiative modes, also known as bound states in the continuum (BICs), are among the most prominent carriers of such singularities, as in the far-field they correspond to points where the polarization state is undefined, and the surrounding region of the momentum space forms a polarization vortex with a well-defined topological charge~\cite{Hsu2016Jul, Koshelev2020Engineering, Zhen, Femius2018}. These vortices underlie topological optical communication schemes~\cite{sinev2025chirality}, vortex lasers with controllable orbital angular momentum~\cite{chen2023observation}, and near-field sensing via local-field enhancement close to the singularity~\cite{lepeshov2023topological}.

For all of these applications, the ability to predict, control and prescribe the topological charge of BIC in advance would offer a decisive advantage. Many recent studies have focused on controlling and increasing the magnitude of the topological charge. In Refs.~\citenum{Li2026, Su2026}, a topological charge of -3 was achieved in periodic structures by merging four off-$\Gamma$ BICs, each with topological charge -1, with a single $\Gamma$-point BIC carrying topological charge +1. Ref.~\citenum{Huanyang2026} demonstrated that the sign of the topological charge can be controlled by inducing its migration between photonic bands through variations in geometric parameters. The authors of Ref.~\citenum{Kang2025} showed that 
topological charges asymmetric with respect to the radiation direction can be generated by engineering the geometric parameters of the structure. 
Ref.~\citenum{Wang2026}
reported a topological charge with an absolute value of 5 in quasi-crystals achieved by employing 
photonic-crystal slabs with different tilt angles; this is the highest topological charge realized in photonic systems to date.

Further advances in the
field of topological-charge control require a rigorous theoretical framework
describing the formation of topological charge. So far, the main theoretical approach is based on the multipolar formalism~\cite{Chen}. Multipolar decomposition is a powerful and well-proven tool for describing BICs. For instance, Ref.~\citenum{Sadrieva} established a connection between the existence of BICs and destructive interference of multipoles. Within the multipolar formalism, the authors of Ref.~\citenum{Chen} proposed associating the topological charge with the Hopf index of the dominant multipole in the mode. They derived a simple expression for the topological charge at the $\Gamma$-point, $q = 1 - |m|$, where $m$ is the azimuthal number of the dominant multipole. However, this approach becomes inapplicable when no single multipole dominates the mode. Moreover, recent work~\cite{Li2026} has reported a case where the topological charge $q=-3$ does not match the index $-1$ of the dominant magnetic multipole with the azimuthal number $m=2$. These inconsistencies suggest that the current approach, while useful, does not yet provide a complete description of topological-charge formation.
This issue calls for a more comprehensive theoretical framework to better understand the origin of topological charge of BICs.

In this Letter, we propose a unified description of
the formation of topological charges, which extends 
to scenarios beyond the $q=1 - |m|$ rule. We provide a dynamical-systems framework for the far-field topology of BICs.
We show that the far-field polarization defines a tangent vector field on the sphere of radiation directions and that the BIC topological charge equals the index of the corresponding singularity. This directly bridges the phase winding of the polarization vortex and the local structure of the vector field near the BIC point. Great attention is devoted to the linearized regime, in which the far-field topology of BICs is determined solely by the Jacobian matrix of the polarization field. Based on the proposed theoretical model, we demonstrate semi-numerically that the asymptotic behavior of the multipoles may have a weaker influence on the Jacobian matrix than the derivatives of the multipolar coefficients. Moreover, our results explicitly show that the index of the dominant multipole may not coincide with the resulting topological charge. These findings reveal the complexity of topological-charge formation and indicate that, in general, the topological charge cannot be reliably obtained from the index of the dominant multipole alone. Furthermore, using a dynamical-systems approach, we demonstrate that nonlinear asymptotic behavior of the polarization field is a necessary condition for realizing high topological charges. We also derive explicit multipolar conditions under which the nonlinear regime can be achieved. This theoretical perspective, which connects differential geometry and dynamical-systems theory with the topological properties of BICs, provides a robust framework for fundamental research and future metasurface design aimed at controlling the magnitude of topological charges.

We begin with the analysis of resonant modes supported by a two-dimensional periodic array of dielectric nanoparticles with arbitrary geometry and material properties. In Ref.~\citenum{Sadrieva}, an explicit expression for the far-field response of modes supported by such metasurfaces was derived. For frequencies below the diffraction limit, the far field can be expressed, up to an overall multiplicative factor, as follows:

\begin{equation}
    \label{eq:1}
    \mathbf{E}(\mathbf{r}) = \mathbf{c}(\mathbf{k}_1) e^{i\mathbf{k}_1\mathbf{r}},
\end{equation}
where $\mathbf{c}(\mathbf{k}_1)$ is the polarization field in the open diffraction channel:

\begin{equation}
\label{polarization}
    \mathbf{c}(\mathbf{k}_1) = 
     \sum_{p_i, p_r, l, m} D_{p_ip_rlm}(\mathbf{k}_1)\mathbf{Y}_{p_ip_rlm}\left(\frac{\mathbf{k}_1}{k_1}\right).
\end{equation}

Here the spherical vector \(\mathbf{Y}_{p_ip_rlm}\left(\mathbf{k}_1/k_1\right)\) is the combined notation for real magnetic \(\mathbf{M}^{e, o}_{lm}\) and electric \(\mathbf{N}^{e, o}_{lm}\) vector spherical harmonics (VSHs). For electric multipole, $p_r=e$ if the multipole is even under reflection from the $y = 0$ plane (polar angle transformation $\varphi \to - \varphi$
in the spherical coordinate system), and $p_r=o$ if the multipole is odd. For magnetic multipoles, the parity is inverse. To distinguish between the magnetic and electric types, spatial inversion parity \(p_i\) is used: \(p_i = (-1)^l\) for electric multipoles, \(p_i = (-1)^{l+1}\) for magnetic multipoles. Index $l$ is the multipole order, and $m$ varies from $0$ to $l$. The symbol \(\mathbf{k_1}\) stands for the wave vector outside the material, with in-plane part \(\mathbf{k_{1||}} = \mathbf{k_b}\), where \(\mathbf{k_b}\) is the Bloch vector. Multipole weights are denoted as \(D_{p_ip_rlm}\).

\begin{figure*}[th!]
\centering
\includegraphics[width = 2\columnwidth]{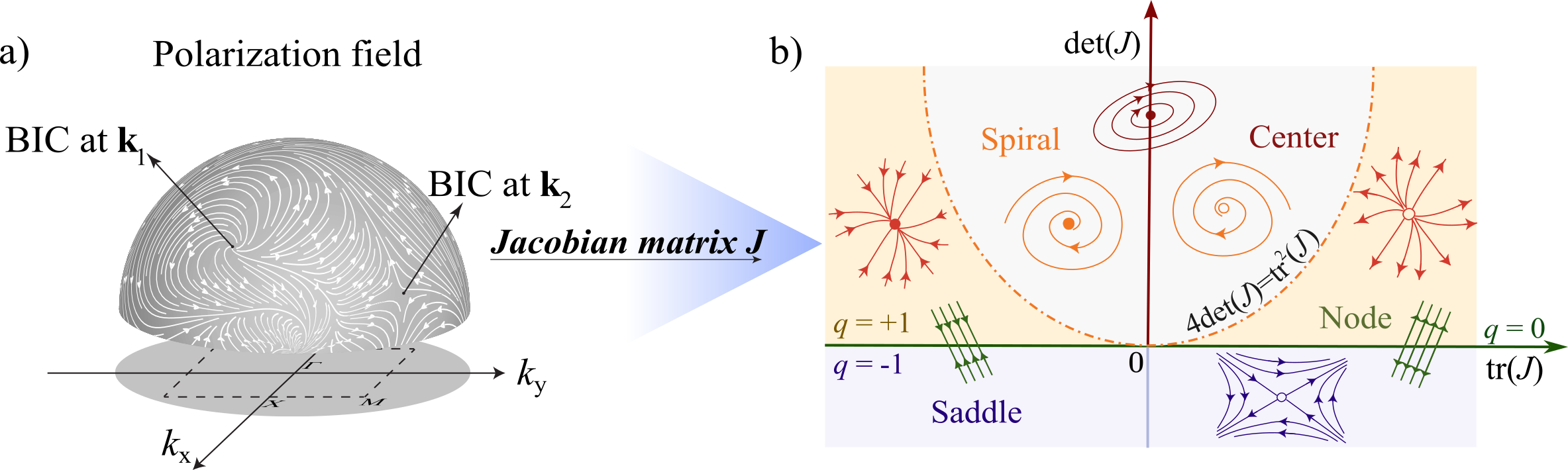}
\caption{(a)
Polarization field of the far‑field radiation on a sphere, defined by the directions of the wave vector outside
the structure. Symbols $\mathbf{k}_1$ and $\mathbf{k}_2$
indicate the BIC points. The first Brillouin zone (of a two‑dimensional C$_4$ lattice, for example) is marked by dashed lines. (b) All possible types of behavior of the field lines of polarization around the BIC point, depending on the trace $\mathrm{tr} J$ and the determinant $\mathrm{det} J$ of the Jacobian matrix $J$ of the polarization field. The indices  (topological charges)  of the BIC points are denoted as $q$.}
\label{linear phase portraits}
\end{figure*}

In this paper, we assume that $\mathbf{c}(\mathbf{k}_1)$ is a purely real vector corresponding to linear polarization. Strictly speaking, the polarization field can be regarded as real in the vicinity of a BIC; therefore, the subsequent analysis remains valid locally even if the polarization field is not strictly real. Otherwise, in the absence of $C$-points, the analysis can be reformulated in terms of the semi-major axis of the polarization ellipse. In the general case, however, a complex polarization field requires a more detailed treatment.

Next, we develop an equivalent mathematical formulation of topological charge formation using the dynamical systems formalism. First, by construction, the polarization field $\mathbf{c}(\mathbf{k}_1)$ is a tangent field to the sphere $S^2$ defined by the unit vector $\mathbf{e}_{\mathbf{k}_1} = \mathbf{k}_1/k_1$. 
It means that $S^2$ includes all directions associated with $\mathbf{k}_1$. Second, it is a fundamental geometric fact that any sphere $S^n$ is a differentiable manifold, especially $S^2$ in our case. Therefore, from the first and second statements, $\mathbf{c} (\mathbf{e}_{\mathbf{k}_1})$ is a vector field on sphere, or a mapping from the manifold to its tangent bundle: $S^2 \to TS^2$. We assume that the field $\mathbf{c} (\mathbf{e}_{\mathbf{k}_1})$ is a non-zero vector in $T_{\mathbf{e}_{\mathbf{k}_1}} S^2$ in the compact area of $S^2$; therefore, we have a
one-parameter group of diffeomorphisms $g^{\gamma}: S^2 \to S^2$, for which the vector field $\mathbf{c} (\mathbf{e}_{\mathbf{k}_1})$ is a field of phase velocity, described by the following equation (see Ref.~\citenum{Arnold1992} for mathematical details):

\begin{equation}
\label{eq:phase velocity}
   \mathbf{c} (g^{\gamma} \mathbf{e}_\mathbf{k_1}) = \dfrac{\mathrm{d}}{\mathrm{d} \gamma} g^{\gamma} \mathbf{e}_\mathbf{k_1}.
\end{equation}

From a physics perspective, Eq.~\eqref{eq:phase velocity} means that 
there are force lines of the polarization field on the surface of the sphere. The dynamic parameter $\gamma$ plays the role of effective "time" that defines motion along the field lines of force. Meanwhile, the BIC point $\mathbf{k}_0$ is related to the stationary point $\mathbf{e}_{\mathbf{k}_0}$ of Eq. \eqref{eq:phase velocity}. In turn, topological charge can be considered as the index of this stationary point: $q=\mathrm{Ind}(\mathbf{e}_{\mathbf{k}_0})$, similarly, but not equivalently, to the concept described in Ref.~\citenum{Chen}. Thus, to reveal the polarization field around the BIC point, it is sufficient to study the phase portrait of the corresponding dynamical system, Eq.~\eqref{eq:phase velocity}. In the consequent steps, we connect the phase portrait analysis and topological charge estimation with the parameters of the polarization field near the stationary points of Eq.~\eqref{eq:phase velocity}.

The first step is linear analysis, which boils down to analysis of Jacobian matrix of polarization field. We expand the polarization field in the vicinity of the BIC point $\mathbf{e}_{\mathbf{k}_0}$ as follows:

\begin{equation}
    \mathbf{c}(\mathbf{e}_{\mathbf{k}_1}) = J(\mathbf{e}_{\mathbf{k}_1} - \mathbf{e}_\mathbf{k_0})
    + o(\mathbf{e}_{\mathbf{k}_1} - \mathbf{e}_\mathbf{k_0}),
\end{equation}

where

\begin{multline}
\label{Jacobian matrix}
    J = \sum_{p_i, p_r, l, m}  \left[ D_{p_i p_r l m} (\mathbf{k}_0) \cdot  \left. \dfrac{\partial \mathbf{Y}_{p_i p_r l m} }{\partial \mathbf{e}_\mathbf{k_1}}\right|_{\mathbf{e}_{\mathbf{k}_1} = \mathbf{e}_{\mathbf{k}_0}}  \right. \\ 
    + \left. \left. \dfrac{\partial D_{p_i p_r l m} }{\partial \mathbf{e}_{\mathbf{k}_1}}\right|_{\mathbf{e}_{\mathbf{k}_1} = \mathbf{e}_{\mathbf{k}_0}} \otimes  \mathbf{Y}_{p_i p_r l m} (\mathbf{k}_0) \right]
\end{multline}

is the Jacobian matrix of the polarization field.

It should be noted that the derivatives $\partial \mathbf{Y}_{p_i p_r l m} / \partial \mathbf{e}_{\mathbf{k}_1}$ are not, in general, well-defined on manifolds (in our case, $S^{2}$), which necessitates the introduction of covariant derivatives. However, in the present analysis, we effectively straighten the field in the vicinity of the BIC point; this procedure is a smooth deformation and therefore preserves the topological properties. Accordingly, we restrict ourselves to a local chart 
in the neighborhood of $\mathbf{e}_{\mathbf{k}_0}$, where ordinary derivatives are well-defined.

If the Jacobian matrix~\eqref{Jacobian matrix} is non-zero (which means all its elements are non-zero), then its determinant $\mathrm{det}J$ fully dictates the allowed values of the topological charge. Direct calculations presented in the Supplementary Material, Sec.II~\cite{suppl}, show that the topological charge is given by $q=\mathrm{sign}(\mathrm{det}J) $. Consequently, in the case of non-zero Jacobian matrix, the absolute value of the topological charge cannot exceed unity.

Moreover, the analysis of the
Jacobian matrix~\eqref{Jacobian matrix} provides a systematic way to identify the possible polarization patterns in the vicinity of the BIC point, which correspond to local phase portraits of the dynamical system~\eqref{eq:phase velocity}. The classification of phase portraits for two-dimensional dynamical systems is wel-established and has been extensively studied in the literature~\cite{Bautin1990}. Here, we briefly summarize the main results relevant to our analysis.

If the Jacobian matrix is nondegenerate (i.e., both of its eigenvalues are nonzero), the local field patterns are fully defined
by the relationship between the trace $\mathrm{tr}J$ and the determinant $\mathrm{det}J$ of the Jacobian matrix~\eqref{Jacobian matrix}. The resulting classification of field patterns as a function of 
$\mathrm{tr}J$ and $\mathrm{det}J$ is summarized in Fig.~\ref{linear phase portraits}. Importantly, the case of $\mathrm{tr}J=0$ and $\mathrm{det}J>0$ does not allow distinguishing between center-type and spiral-type phase portraits without additional analysis of the vector field, in accordance with the Grobman–Hartman theorem~\cite{Grobman}. In contrast, if the Jacobian matrix is degenerate (i.e., at least one eigenvalue vanishes, which is equivalent to $\mathrm{tr}J=0$), the linear approximation becomes insufficient, and nonlinear analysis is required, extending beyond the immediate vicinity of the singular point. Nevertheless, the corresponding local phase portraits are also illustrated in Fig.~\ref{linear phase portraits}.

\begin{figure*}[t!]
    \centering
    \includegraphics[width=\linewidth]{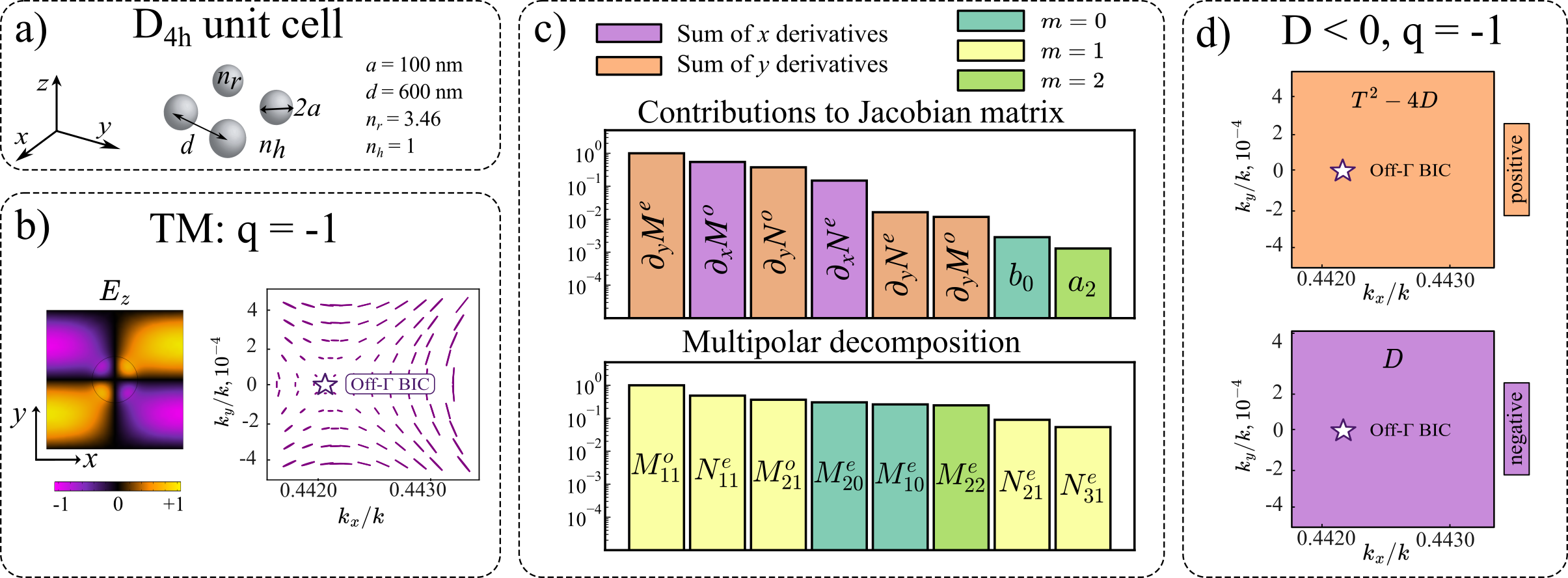}
    \caption{(a) Schematic illustration of a unit cell with D$_{4h}$ symmetry corresponding to a metasurface composed of dielectric spheres; the material and geometric parameters are indicated. (b) Left: normalized field distribution of a BIC in the $\Gamma X$-valley at $k_{0_x} = 0.54\pi/d$. Right: far-field polarization field distribution in the reciprocal space: it ce 
    exhibits a saddle-type phase portrait with the corresponding topological charge $q=-1$. (c) Multipolar decomposition of BIC`s far-field and the resulting contributions to the Jacobian matrix. The derivatives of the multipolar coefficients provide a dominant contribution compared to that of the multipolar asymptotics. (d) Sign distributions of $\mathrm{det} J$ and  $\mathrm{tr}^2 J -4 \mathrm{det} J$ in the vicinity of the BIC, where we introduced notations $D = \mathrm{det} J$ and $T=\mathrm{tr} J$ for brevity. The signs of these quantities are in full agreement with the observed topological charge and the type of polarization field.}
    \label{fig:off-gamma_num}
\end{figure*}

We numerically validate the proposed analytical model using finite elements method (FEM) simulations. As a representative example, we consider an off-$\Gamma$ BIC supported by a square array of dielectric spheres with parameters specified in Fig.~\ref{fig:off-gamma_num}. The far-field polarization distribution in this case exhibits a topological charge of $-1$ and a saddle-type phase portrait. The case of $\Gamma$-BICs is performed and discussed in the End Matter.

We perform a multipolar decomposition of the BIC`s far field in the VSH basis aligned with the BIC direction in the reciprocal space. This rotation of the basis renders the multipolar description effectively equivalent to that at the 
$\Gamma$-point and improves the numerical accuracy. The decomposition reveals that no single multipole dominates the response by several orders of magnitude; therefore, the conventional association of the topological charge with the index of a dominant multipole is not applicable in this case. However,  even if we identify the topological charge with the index of the formally strongest contributing multipole $\mathbf{M}_{11}$ in the decomposition, these two values
do not coincide: the index is 0, whereas the topological charge is -1. A similar discrepancy between the index of the dominant multipole and the topological charge was reported in Ref.~\cite{Li2026}.

Accordingly, within our proposed framework, we compute the Jacobian matrix of the polarization field to account for the observed polarization pattern and the corresponding topological charge. Figure \ref{fig:off-gamma_num}c presents the semi-analytically derived contributions to the Jacobian
matrix, where the following terms are introduced:

\begin{equation}
\begin{split}
\label{eq:determinant}
&J_{11} =  -a_0 + a_2 -\partial_xM^e-\partial_xN^o,\\
&J_{12} = -b_0 + b_2 -\partial_yM^e-\partial_yN^o, \\
&J_{21} = b_0 + b_2 +\partial_xM^o - \partial_xN^e, \\
&J_{22} =  -a_0 -a_2  +\partial_yM^o-\partial_yN^e,
\end{split}
\end{equation}
where we introduced notations

\begin{align}
    &\begin{bmatrix}
a_0 \\
b_0
\end{bmatrix} = \sum_{l=1}^\infty D^{\begin{bmatrix}
N \\
M
\end{bmatrix} e}_{0l}\frac{l(l+1)}{2}, \\
    &\begin{bmatrix}
a_2 \\
b_2
\end{bmatrix} = \sum_{l=2}^\infty (D^{N \begin{bmatrix}
e \\
o
\end{bmatrix}}_{2l} \pm D^{M \begin{bmatrix}
o \\
e
\end{bmatrix}}_{2l})\frac{(l+2)!}{4(l-2)!},
\end{align}
which correspond to the impact of the multipoles' asymptotics, and
\begin{align}
    &\partial_sA^{p_r} = \sum_{l=1}^\infty\frac{\partial D^{
    A}_{p_rl1}}{\partial s}\dfrac{l(l+1)}{2}, & A = \{M, N\}, s =\{x, y\}, 
\end{align}
which correspond to the impact of multipolar coefficients'
derivatives along the $x$ and $y$ directions associated with the corresponding Cartesian coordinate system of the tangent plane to the sphere defined by normalized wave vector at the BIC point. 
This impact has never been considered before.

Figure~\ref{fig:off-gamma_num}(c) demonstrates the dominant impact of multipolar coefficients'
derivatives rather than the impact of multipoles'
asymptotics; this behavior has not been reported previously. Based on the computed Jacobian matrix, we evaluate the quantities $\mathrm{tr}^2 J - 4 \mathrm{det} J$ and $\mathrm{det} J$ both at the BIC point and in its vicinity, to account for numerical inaccuracies associated with lack of the exact BIC position. The obtained results, $\mathrm{tr}^2 J - 4 \mathrm{det} J>0$ and $\mathrm{det} J < 0$, are in full agreement with the numerically observed saddle-type polarization pattern and the corresponding topological charge $q=-1$.

As noted above, when the Jacobian matrix of the polarization field is non-zero, the absolute value
of the topological charge cannot exceed unity. In contrast, a vanishing Jacobian matrix allows for higher topological charges. To describe this regime, we employ results from dynamical systems theory. Specifically, if the polarization field exhibits an $n$-th degree polynomial asymptotic behavior as a function of direction in the reciprocal space in the vicinity of the BIC, then the topological charge satisfies the estimate $|q| \leq \Pi_2(n+1)=n$, where $\Pi_2(n+1)$
denotes the Petrovskii number~\cite{Arnold1978}. At the same time, in the nonlinear regime, there is no general classification of the exact indices of singular points or corresponding phase portraits. Nevertheless, specific forms of the polarization field can be constructed that yield higher topological charges, as discussed in the Supplemental Material, Sec.VII~\cite{suppl}.

There are two possible scenarios in which the Jacobian matrix of the polarization field vanishes: (i) all the terms in Eq.~\eqref{Jacobian matrix} vanish individually at the BIC point $\mathbf{e}_{\mathbf{k}_0}$; or (ii) nonzero terms cancel each other, resulting in a zero sum. Scenario (i) requires both the multipoles' amplitudes and their Jacobian matrices to vanish at the BIC point

At the poles of the radiation-direction sphere, 
which correspond to the $\Gamma$ point, the multipoles with $m=0$ and $m=2$ vanish, whereas their Jacobian matrices are nonzero. Conversely, the multipoles with $m=1$ have nonzero values, while their Jacobian matrices vanish. For $m \geq 3$, both the multipole amplitudes and their Jacobian matrices vanish. Therefore, realizing the nonlinear regime through scenario (i) requires suppressing the multipoles with $m=0$ and $m=2$ and eliminating the contributions from the derivatives of the multipolar coefficients for $m=1$. The latter condition can be achieved, for example, in the presence of flat bands, since $\partial D_{p_i p_r l m}/\partial \mathbf{e}_{\mathbf{k}_1} = \partial D_{p_i p_r l m}/\partial \omega \cdot \partial \omega / \partial \mathbf{e}_{\mathbf{k}_1}$, or through symmetry constraints, as discussed in the End Matter. Away from the poles, along the directions
$(\theta_0=\pi/2;\phi_0 = 0, \pi, \pi/m)$ ,
odd multipoles with even $l-m$, as well as even multipoles with odd $l-m$, have vanishing Jacobian matrices while remaining nonzero. Similarly, along the directions $(\theta_0=\pi/2;\phi_0 = \pi/2, \pi/2m)$,
odd multipoles with odd $l-m$, as well as even multipoles with even $l-m$, have vanishing Jacobian matrices while remaining nonzero. In these cases, the nonlinear regime under scenario (i) can be obtained only if the derivatives of the corresponding multipolar coefficients also vanish. All other points in the reciprocal space can support the nonlinear regime only through scenario (ii), in which nonzero contributions to the Jacobian matrix cancel one another. By tuning the geometric and material parameters of the system, one can modify the multipolar coefficients and the multipolar composition of the mode, thereby controlling the polarization-field pattern and the value of the topological charge through either scenario (i) or scenario (ii).

In summary, we have developed a theory describing the formation of topological charges BICs.
We show that the emergence of polarization patterns and topological charges is governed by the phase portraits and indices of the stationary points of the dynamical system formed by polarization-field force lines
on the two‑dimensional sphere of radiation directions (or, locally, $\mathbb{R}^2$). The theoretical description is validated numerically in the linear regime. Our simulations consider
cases in which no single multipole dominates the mode, rendering the traditional identification of topological charge with the index of a dominant multipole inapplicable. In contrast, the proposed theory fully explains the numerically obtained topological charge and the observed polarization-field pattern. Furthermore, the numerical results reveal a strong influence of the derivatives of multipolar coefficients on the topology of BICs—an effect not described previously. 
On theoretical grounds, we argue that the formation of large topological charges is associated with the vanishing of polarization-field derivatives, which leads to nonlinear asymptotics of the polarization field in the reciprocal space near the BIC. Possible ways of realizing
such nonlinear asymptotics, including symmetry considerations, were discussed. Our future work will study the nonlinear regime in detail. The approach and results presented here offer practical means to control polarization-field patterns and topological charges by tuning the geometric and material parameters of photonic systems. Moreover, the established link between differential geometry, dynamical-systems theory, and the topology of BICs can be extended to other optical singularities, providing a powerful framework for describing and controlling optical topological properties.

\section*{Acknowledgments}

Analytical studies were supported by Russian Science Foundation grant № 26-72-10063. Numerical studies were supported by Russian Science Foundation grant № 25-42-10025. The authors thank Lydia Pogorelskaya for proofreading the manuscript. The authors also thank Andrey Bogdanov for fruitful discussions.
\bibliography{Bibliography1}

\appendix 
\onecolumngrid

\section*{End matter}

For \(\Gamma\)-BIC, we can 
leverage the symmetry of the structure to restrict the impact of possible multipolar coefficients and their derivatives to the Jacobian matrix, which affects the possible values of topological charges and phase portraits. Let 
permittivity be invariant under the \(C_n\) rotational group \((n>2
)\). This group has $n$ irreducible representations, 
and we assume the mode is transformed through the $j$th irreducible representation. 





First, 
it is well-known that 
the mode in the $j$th representation has nonzero coefficients only for multipoles with \(m = j+tn, t\in \mathbb{N}\). Second, we explicitly showed in Supplemental Materials VI~\cite{suppl} that 
first derivatives of the multipolar coefficients are not zero only if \(j-m = \pm 1\).

In linear expansion, we care only for coefficients with \(m=0, 2\) and derivatives with \(m = \pm 1\), so the Jacobian matrix will not be trivial only in the $0$th and $2$nd representations. 
In our notations, 
\(m = \pm1\) derivatives along the $x$ and $y$ 
directions  are related as follows:

\begin{equation}
    \partial_yN^e = -i^j\partial_xN^o, \quad \partial_yN^o = i^j\partial_x N^e, \quad \partial_yM^e = -i^j \partial_xM^0,  \quad \partial_y M^o = i^j \partial_x M^e.
\end{equation}

As a result, for representation $j=0$, we 
obtain the following conditions on the trace and determinant of the Jacobian matrix:
\begin{equation}
\begin{split}
    &\mathrm{tr}J= -2(a_0+\partial_xM^e+\partial_xN^o),\\
    &\det J = (a_0+\partial_xM^e+\partial_xN^o)^2+( b_0 - \partial_xN^e+\partial_xM^o)^2 > 0, \\
    &\mathrm{tr}^2 J-4\det J = -4( b_0 - \partial_xN^e+\partial_xM^o)^2\leq 0,
\end{split}
\end{equation}
which leads to topological charge $q=1$, while the phase portrait cannot be exactly determined without knowing additional reflection symmetry, which 
we discuss further.

For the representation $j=2$, we have:
\begin{equation}
\begin{split}
    &\mathrm{tr} J = 0,\\
    &\det J = -(a_2-\partial_xM^e-\partial_xN^o)^2-(b_2-\partial_xN^e+\partial_xM^o)^2 < 0,
\end{split}
\end{equation}
which leads to topological charge $q=-1$ and a saddle-type phase portrait.


\begin{figure*}[t!]
    \centering
    \includegraphics[width=0.9\linewidth]{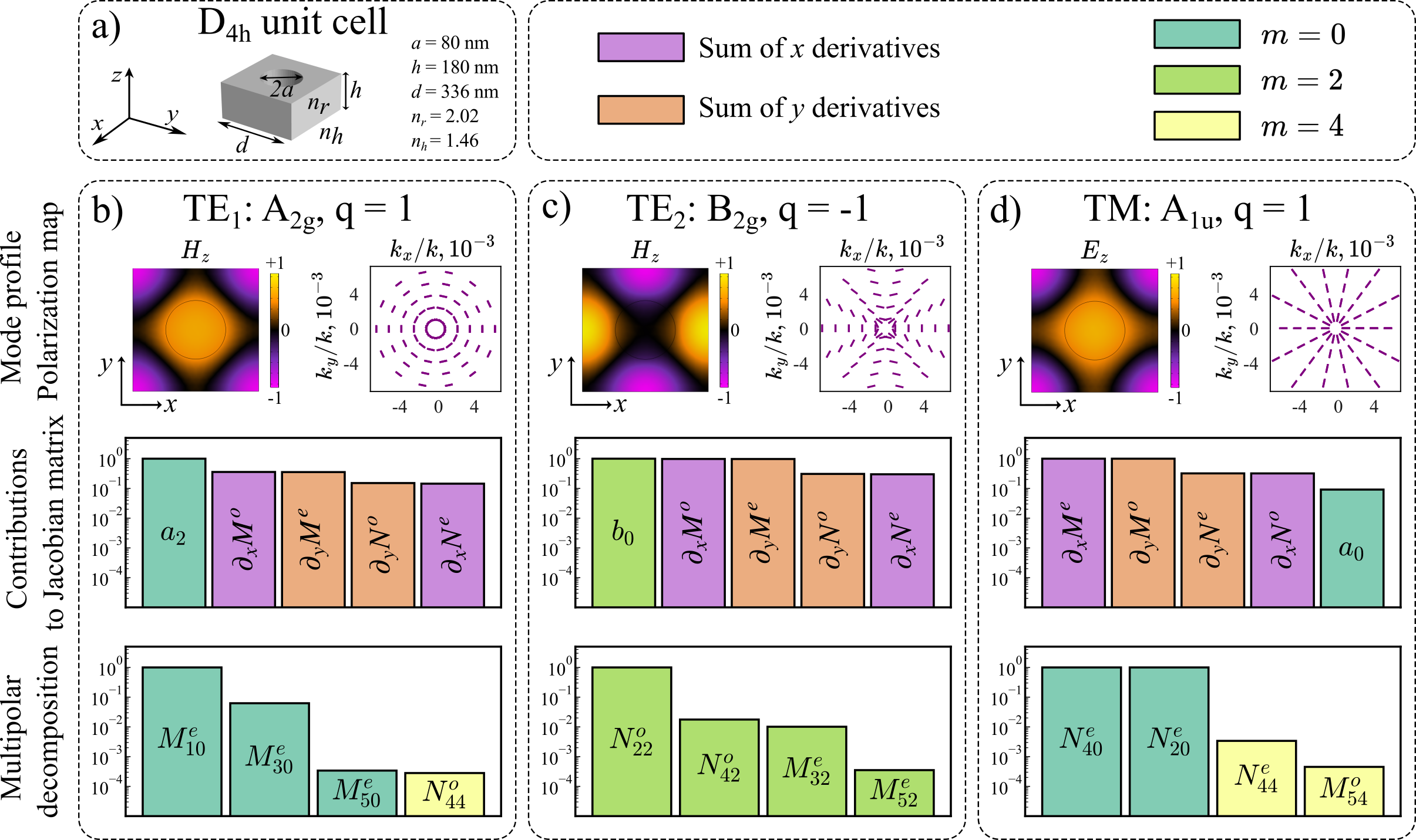}
    \caption{(a) Schematic illustration of a unit cell with D$_{4h}$ symmetry corresponding to a photonic crystal with cylindrical holes; the material and geometric parameters are indicated. (b) - (d) Mode profiles, polarization maps, Jacobian-matrix contributions, and multipolar decompositions for representative \(\Gamma\)-BICs with center, saddle, and node phase portraits. As predicted in theory, the representations A ($j = 0$) and B ($j = 2$) have topological charges $q = +1$ and $q = -1$, respectively. The multipolar content follows $m=j+4p,p\in \mathbb{N}$, two pairs of equal x and y derivatives are present and other four derivatives are zero due to the additional reflection symmetry $y\rightarrow-y$.}
    \label{fig:gamma}
\end{figure*}

Next, we numerically  
verify our theoretical group predictions. We find  
three $\Gamma$-BICs in the structure with the unit cell presented in 
Fig.~\ref{fig:gamma}(a). These three 
modes give 
three phase portraits corresponding to the non-zero Jacobian matrix: TM\(_1\) - center, TM\(_2\) - saddle, and TE - node (see Fig.~\ref{fig:gamma}(b, c, d)). Mode profiles and polarization maps 
reconstructed from multipolar coefficients and multipolar contents at the $\Gamma$ point are presented in 
Fig.~\ref{fig:gamma}(b, c, d). We have obtained the same results as shown in theory: if the mode corresponds to the $0$th representation (as TM\(_1\) and TE modes), \(q = 1\); if the mode is in the $2$nd representation (as TM\(_2\) mode), \(q = -1\). The multipolar content matches the representation: \(m = j+4p, p\in \mathbb{N}\), and the required derivative relations are fulfilled. Due to extra reflection symmetry \(y\rightarrow-y\), \(a_0 = \partial_xM^e = \partial_xN^o = 0\) for the TM\(_1\) mode,  which yields \(\mathrm{tr}J = 0\) and results in a center phase portrait, consistent with the theoretical predictions. For the TE mode, reflection symmetry dictates
\(b_0 = \partial_xN^e = \partial_xM^o = 0\), giving \(\mathrm{tr}^2J -4\mathrm{det} J = 0\) and a node phase portrait, also in agreement with theory. 

Thus, we showed how symmetry of the unit cell may affect different contributions to Jacobian matrix resulting in the pattern of polarization field near BIC point and the value of topological charge. Our proposed approach is consistent with results  reported in previous papers \cite{Arjas2024Nov} 
in which the topological charge is fully determined by the irreducible representation of the rotational symmetry group associated with the mode. In addition, we demonstrate that the full symmetry group further constrains the set of permitted
patterns of the polarization field near BIC point.

\end{document}